\documentclass[aps,prl,twocolumn,groupedaddress,showpacs]{revtex4-1}
\usepackage{graphicx}
\usepackage{amsmath}
\begin{document}

%\preprint{}

\title{Accurate feeding of nano antenna by polarization singularities for lateral and rotational displacement sensing}


\author{Zheng Xi}
\email[z.xi@tudelft.nl]{}
\affiliation{Optica Reseach Group, Delft University of Technology}
\author{Lei Wei}
\affiliation{Optica Reseach Group, Delft University of Technology}
\author{A.J.L. Adam}
\affiliation{Optica Reseach Group, Delft University of Technology}
\author{H.P. Urbach}
%\thanks{}
\affiliation{Optica Reseach Group, Delft University of Technology}


\date{\today}

\begin{abstract}
Addressing subwavelength object and displacement is crucial in optical nanometrology. We show in this Letter that nano antennas with subwavelength structures can be addressed precisely by incident beams with singularity. This accurate feeding beyond the diffraction limit can lead to dynamic control of the unidirectional scattering in the far field. The combination of polarization discontinuity of the incoming singular beam, along with the rapid phase variation near the antenna leads to remarkable sensitivity of the far field scattering to displacement at deep subwavelength scale. This opens a far field deep subwavelength postion detection method based on the interaction of singular optics with nano antennas.
\end{abstract}
%\keywords{}
\maketitle
%\section{}
Great efficiencies have been achieved using nano antenna to control light at the nanoscale\cite{novotny2011antennas}\cite{bharadwaj2009optical}. The recently developed antenna array at optical frequency domain, also called metasurface, has propelled tremendous applications of controlling the direction of far field scattering\cite{bliokh2015spin,lin2013polarization,yu2011light,yu2014flat}. The dynamic control over the scattering directivity requires precise excitation of the appropriate currents in the antenna\cite{Xi:16}. Since the building blocks of ultracompact nano antennas are subwavelength nanoparticles whose size and inter-particle gaps are both beyond the diffraction limit, it is in general quite challenging to precisely feed these arrays using far field optical schemes. Previous works addressed this issue using polarization selective excitation of different antenna elements to realize directional scattering\cite{lin2013polarization,lindfors2016imaging}. These antenna arrays are based on similar concept as for those used in microwave antenna arrays, the gaps between individual element are comparable to the working wavelength. One of the promising approaches is to use spatially inhomogeneous light to control the near field hotspot produced by the nano antennas\cite{volpe2009controlling}. However, it still remains an open question how to precisely address on deep subwavelength antennas for the active control of the far field unidirectional scattering. 

Singular optics has been an intriguing candidate for the study of techniques far beyond the diffraction limit\cite{dennis2009singular}. It usually refers to discontinuities in the light field itself, normally the polarization singularity and phase singularity where the field at a singular point or line vanishes.  The non-trivial  behavior of the extremely fast phase change at the singularity has led to important applications such as high-precision nanoscale metrology as well as super resolution imaging\cite{hell2009diffraction,fischer2013three,shamir2012singular,roy2015radially,neugebauer2015polarization,rodriguez2010optical,neugebauer2015measuring,neugebauer2014polarization,bauer2014nanointerferometric,volpe2009controlling}. 

In this Letter, we use the polarization singularity at the beam centre to precisely address one individual element of two identical parallel metallic nanorods seperated by a deep subwavelength gap. The precise alignment of the singular point with the two nanorods gives rise to an accurate feeding of the nano antenna far beyond the diffraction limit.  This results in the dynamic control of unidirectional scattering in the far field. Because a singularity is a point, it is possible to achieve very high resolution in the detection of the lateral and rotational displacement using proper combination of singular beams and resonant nano antennas. We show results obtained by rigorous simulations and present a sufficiently accurate model based on two interacting dipoles to explain the rapid change of scattering pattern within subwavelength displacement. In contrast to common understanding, in the proposed scheme, it is better to use longer wavelength for the detection of smaller displacement. The detection sensitivity can be increased by forming arrays of these nano antennas.

\begin{figure}

%\fbox
\includegraphics[width=\linewidth]{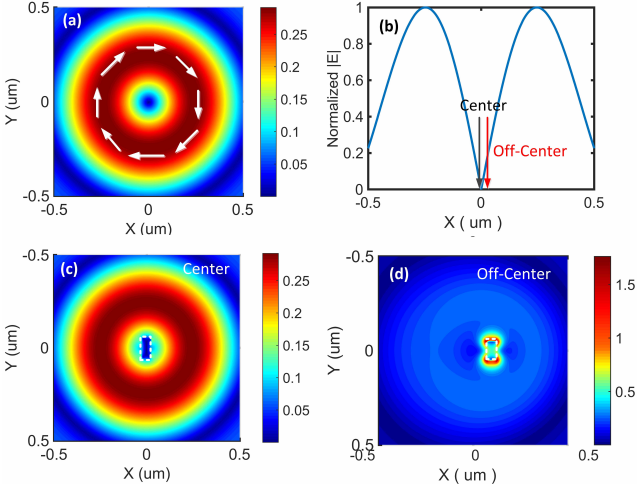}
\caption{(a) Electric field intensity $|E|$ distribution at the focal plane of a focused azimuthally polarized beam. The white arrows show the direction of polarization. (b) A linecut across the centre of the beam showing the absolute zero singularity point. (c) Electric field  intensity $|E|$ distribution when a resonant nanorod antenna is placed at the centre singularity point of the azimuthally polarized beam. (d) Electric field intensity $|E|$ distribution when a resonant nanorod antenna is displaced 100nm from the centre singularity point of the azimuthally polarized beam. The antenna is a gold nanorod antenna with dimensions 50nm x 50nm x 100nm as shown as the doted rectangular area. The wavelength is 791nm.}
\label{fig:false-color}
\end{figure}

Let us first consider the case of a focused azimuthally polarized beam shown in Fig. 1(a). The numerical aperture is chosen as NA=1.45 with the surrounding n=1.5 throughout this Letter. The focused beam is calculated through Richard-Wolf diffraction integral and implemented into a finite difference time domain simulation. The azimuthal polarization in the pupil causes a polarization singularity at the centre of the focused spot around which the E field is polarized azimuthally and varies very rapidly (Fig. 1b). Consider a single gold nanorod. If the singularity of the illuminating beam is at the centre of the nanorod as shown in Fig. 1(c), due to the symmetry of the field, its dipole moment cannot be excited efficiently and this results in very low field intensity around the antenna. However, if the nanorod antenna is displaced slightly, the broken symmetry induces a resonant dipole moment inside the antenna as shown in Fig. 1(d). This position dependent excitation of the antenna provides a control mechanism of the amount and the directivity of the scattered radiation and forms the basis for drastically change of the far field pattern when two of such antennas are considered.

\begin{figure}

%\fbox
\includegraphics[width=\linewidth]{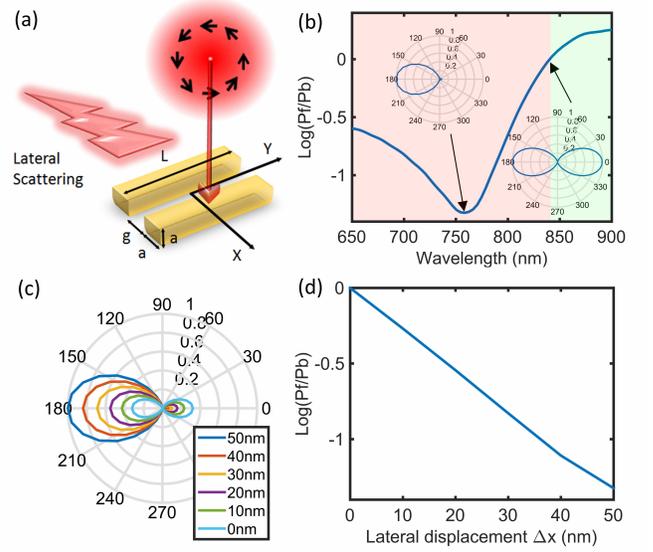}
\caption{(a) Schematic illustration of the considered configuration, the origin of the coordinate system is chosen at the centre of the two antennas. The antenna is made of two nanorods with dimensions a=50nm, g=50nm and L=100nm. (b) The ratio of the front ($x>0$) to back ($x<0$) scattered power. The insets show the scattering patterns in the XY plane at 758 nm and 840 nm, respectively. (c) Change of scattering pattern as the singularity point is displaced from the origin at 758nm. (d) Change of front to back power ratio as the singularity point is displaced from the origin at the wavelength of 758nm.}
\label{fig:false-color}
\end{figure}

In Fig. 2, we study the configuration when two identical gold nanorods are placed parallel to each other and the azimuthally polarized beam is focused at the centre of one nanorod shown in Fig. 2(a). The two nanorods are of dimensions a=50nm, L=100nm and the gap size is g=50nm. The front to back ratio defined as the far field scattered power in the +x direction divided by that in the $-x$ direction is plotted in Fig. 2(b). A logarithmic scale is used for better illustration. The green and red shaded areas are wavelength range where the antenna scatters more in the +x forward or $-x$ backward direction. It is seen that the scattering pattern is very sensitive to wavelength. At 758nm, the antenna at the centre of the focused beam works as a nearly perfect reflector, reflecting most of the energy to $-x$ direction whereas at wavelength longer than 840nm, this centre antenna directs more power in the +x direction. It is interesting to note that at 840nm, although the asymmetry exists both at the illumination and the antenna structure, the far field scattering pattern is highly symmetric. More surprisingly, when the beam is scanned through the antenna for 50nm along the $-x$ axis as shown in Fig. 2(c), which is about $0.066\lambda_{0}$ where $\lambda_{0}$ is the free space wavelength, the scattering pattern changes dramatically from highly asymmetrical unidirectional scattering to a symmetrical scattering pattern. In Fig. 2(d), the power ratio changes from -1.3 to 0 as the singularity point is displaced from 50nm to 0nm from the origin. This drastic change of the far field scattering pattern results in remarkable sensitivity to subwavelength displacement.

A semi-analytical theory illustrated in Fig. 3(a) can be applied to explain this phenomenon. The two nanorod antennas are approximated by two point dipoles $\vec{p}_{1}$ and $\vec{p}_{2}$  placed at a centre-to-centre distance d apart from each other. The phase difference between the two dipoles is $\Delta\Phi$. The incident fields $E_1$ and $E_2$ at the centres of the two nanorods depend strongly on the lateral displacement. Consider the coupling between the two dipoles, with dipole vectors $\vec{p}_1$ and $\vec{p}_2$:

\begin{equation}
\begin{aligned}
  \vec{p}_{1} &= \epsilon_{0} \epsilon_{r} \alpha \vec{E}_{1} + \epsilon_{0} \epsilon_{r} \alpha E_{n} \vec{p}_{2}  \\
  \vec{p}_{2} &= \epsilon_{0} \epsilon_{r} \alpha \vec{E}_{2} + \epsilon_{0} \epsilon_{r} \alpha E_{n} \vec{p}_{1}  \\
\end{aligned} 
\end{equation}
with $\alpha$ being the polarizability of each nanorod and $E_n$ being the near field coupling term:

\begin{equation}
  E_n = - e^{ikd} \frac{1-ikd-k^2d^2}{4 \pi d^3 \epsilon_{0} \epsilon_{r}}
\end{equation}

The total far-field scattering pattern of the two antennas can be calculated using the far field expressions for radiation of two dipoles. The power ratio in the forward and backward direction is chosen as a criteria for evaluating the far-field performance\cite{vercruysse2013unidirectional,xi2013tailoring}
\begin{equation}
{{P}_{f}}/{{P}_{b}}=\frac{|{{p}_{1}}{{|}^{2}}+|{{p}_{2}}{{|}^{2}}+2|{{p}_{1}}||{{p}_{2}}|\cos (\Delta \phi -kd)}{|{{p}_{1}}{{|}^{2}}+|{{p}_{2}}{{|}^{2}}+2|{{p}_{1}}||{{p}_{2}}|\cos (\Delta \phi +kd)}   
\label{eq1}
\end{equation}           

When the singularity point is focused at the centre of the two antennas, $\vec{E}_1 = -\vec{E}_2$ because of the anti-symmetric distribution in the illumination polarization. The two induced dipole moments are:

\begin{equation}
\begin{aligned}
   \vec{p}_{1} &= \epsilon_{0} \epsilon_{r}\frac{\alpha - \alpha^2 E_n}{1-\alpha^2 E_n^2} \vec{E}_{1} \\
	 \vec{p}_{2} &=- \epsilon_{0} \epsilon_{r} \frac{\alpha - \alpha^2 E_n}{1-\alpha^2 E_n^2} \vec{E}_{1} 
\end{aligned}
\end{equation}
The anti-symmetric distribution of polarization in the illumination with respect to the singularity centre is passed to the phase difference of the two induced dipoles inside the two antennas. The two dipoles are of equal strength but of a $\pi$ phase difference. According to Eq. 3, the power in the forward and backward direction is the same and the scattering pattern is symmetric.                                                                      

The situation begins to change drastically if the singularity point is focused at the centre of one of the antenna as shown schematically in Fig. 2(a). In this case, the excitation field $\vec{E}_2=0$, we get:

\begin{equation}
\begin{aligned}
   \vec{p}_{1} &=\epsilon_{0} \epsilon_{r}\frac{\alpha}{1-\alpha^2 E_n^2} \vec{E}_{1} \\
	 \vec{p}_{2} &=\epsilon_{0} \epsilon_{r}\frac{\alpha^2 E_n}{1-\alpha^2 E_n^2} \vec{E}_{1} 
\end{aligned}
\end{equation}

The antenna $\vec{p}_2$ is hardly excited by the focused beam, it can only be excited by the near field coupling of the adjacent antenna $\vec{p}_1$. The phase difference between two antennas is no longer determined by the incident beam but instead it is determined by the near field interaction\cite{rolly2011crucial}.

\begin{equation}
\begin{split}
{{\vec{p}}_{2}} &= \alpha E_n \vec{p}_{1} \\
                &=-\alpha {{e}^{ikd}}(1-ikd-{{k}^{2}}{{d}^{2}})/(4\pi \epsilon_{r} {{\epsilon }_{0}}{{d}^{3}}){{\vec{p}}_{1}}
\label{eq2}
\end{split}
\end{equation} 

At very short distance d, the induced dipole $\vec{p}_{2}$ in the centre nanorod varies extremely sensitively to distance as $d^{-3}$. To determine the phase difference $\Delta\Phi$ , the polarizability $\alpha$ is needed. We performed numerical simulations to extract this parameter by calculating the scattered field of a single nanorod by plane wave excitation and relating the scattered field amplitude in the far-field to the radiated field of the induced dipole inside the antenna\cite{poutrina2014multipole}. Suppose the long axis of the nanorod is in y direction. The plane wave is incident along +x direction, with E field polarized along y direction. The scattered electric far field in +x direction is completely y polarized 
\begin{equation}
{{E}_{scatt,y}}=\frac{{{k}^{2}}{{e}^{ikr}}{{p}_{y}}}{4\pi r{{\epsilon }_{0}}{{\epsilon }_{r}}} 
\end{equation}
while the induced dipole moment in the nanorod is 
\begin{equation}
{{p}_{y}}={{\epsilon }_{0}}{{\epsilon }_{r}}\alpha {{E}_{inc,y}}
\end{equation}
Combing these two equations, we arrive at the final express for the polarizability for a single nanorod:

\begin{equation}
\alpha =\frac{4\pi r{{E}_{scatty}}}{{{k}^{2}}{{e}^{ikr}}{{E}_{inc,y}}}
\label{eq3}
\end{equation}
This procedure is applied for every wavelength of interest separately so that the material dispersion of the gold nanorod is taken account of. In Fig. 3(a), we plot the amplitude ratio between the two induced dipoles $\vec{p}_{2}$ and $\vec{p}_{1}$ inside the two nanorods and the phase difference $\Delta\Phi$ between them using Eq. (6). A clear resonant enhancement of $\vec{p}_{2}$ due to plasmonic resonance can be seen. It is important to note that due to the near field coupling between the two nanorods, the two induced dipole moments are not equal. Therefore, the condition for maximum front to back ratio by setting $\Delta \Phi =\pi /2$ when $|{p}_{1}|=|{p}_{2}|$ as indicated by Eq. (3) is no longer valid. In Fig. 3(b), we plot the front to back power ratio calculated using this semi-analytical method. Good agreement is achieved with full numerical simulation shown in Fig. 2(b). The mismatch is due to the change of electric polarizability of individual antenna due to strong near field coupling.  

\begin{figure}

%\fbox
\includegraphics[width=\linewidth]{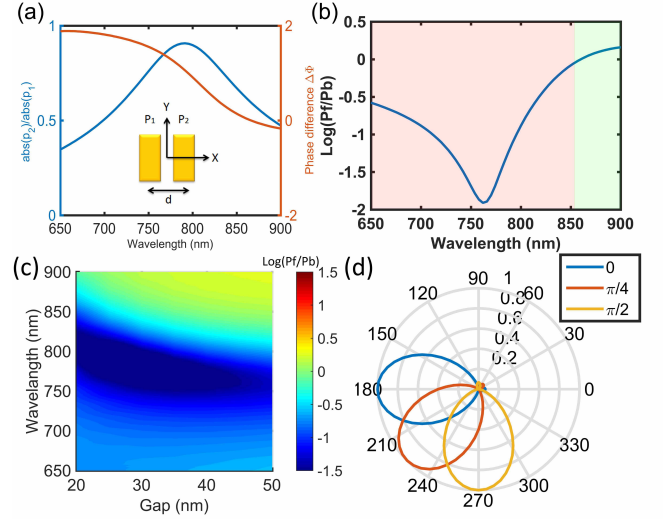}
\caption{(a) Ratio and phase difference of the two induced dipoles calculated using semi-analytical method. Inset shows the model used to describe the unidirectional scattering behaviour. The off-centre antenna is approximated as an electric dipole. Because of the phase asymmetry in forward and backward directions, the electric field interferences constructively in one direction and destructively in the opposite. (b) Front to back ratio obtained through semi-analytical model. (c) Influence on the front to back ratio when smaller gaps ranging from 20nm to 50nm are used. (d) Rotation of the far field scattering pattern as the antenna arrays are rotated over the indicated angles.}
\label{fig:false-color}
\end{figure}

From the above discussion, it follows that the singular point in the azimuthally polarized beam provides a way to accurately control the separate excitations of the two antennas, even when the gap is deep subwavelength. The near field coupling of the two antennas increases rapidly as ${{d}^{-3}}$, providing the required phase difference for unidirectional scattering. We now consider what the effect is of using smaller gaps. Therefore we decrease the gap size g between the two nanorods from 50nm to 20nm and rigorously compute the front to back ratio in Fig. 3(c). A very strong directivity at 800nm can still be observed even when the gap g is only 20nm. Therefore, even for a displacement of 35nm, which is only about $0.04{{\lambda }_{0}}$, the change from a highly asymmetric pattern to a highly symmetric one can still be easily observed. One interesting thing to note here is that as the gap d between the two antenna decreases, the optimal wavelength at which the front to back ratio is maximum shifts to longer wavelength. In contrast with one may expect, in this case a longer wavelength is better suited for the detection of smaller displacements. The explanation for this surprising phenomenon is the plasmonic resonance of the two antennas: The phase difference of the two excited dipole moments is determined by the resonant excitation of the two antennas. As the two antennas approach to each other, their resonance shifts to longer wavelength due to near field coupling, therefore the optimal wavelength for obtaining the best phase difference shifts to longer wavelength. However, although we quantify the displacement in terms of wavelength here, this is actually not a good way of defining the minimum displacement. As we showed above, it is the abrupt polarization change at the beam centre along with the rapid change of phase in the near field that lead to the very high sensitivity of the far field scattering symmetry. The latter depends highly on the resonant coupling of the two antennas. The resonance can be shifted to even longer wavelength when the length of the two antennas is increased while the detection limit is still governed by the gap size plus half width of the antenna. The width of the antenna can also be decreased to move the resonance to longer wavelength. Therefore, the gap size is a more important limiting factor for the displacement detection than the wavelength used.

When the two antennas are rotated with respect to the singularity centre, this rotation in the near field can also be observed by measuring the rotation of the direction of the maximum of the far field scattered power as shown in Fig. 3(d). Because the illumination is rotational symmetric, as the rotation angle changes from 0 to $\pi/2$, the far field scattering pattern changes accordingly. This offers a highly sensitive method of detecting small rotations.

\begin{figure}
%
%%\fbox
\includegraphics[width=\linewidth]{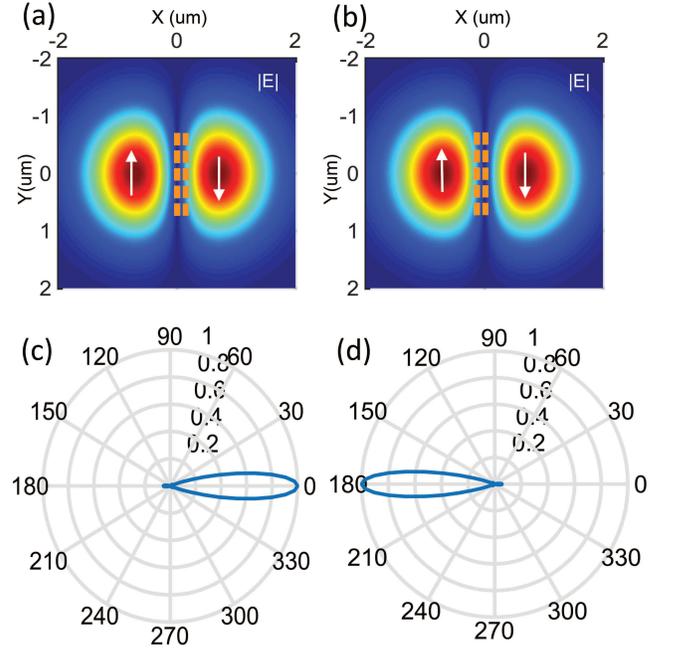}
\caption{(a),(b) Configuration of precise alignment of the antenna arrays with respect to Hermite-Gaussian beam with a dark singularity line in the middle. The polarization is along y direction indicated by the white arrow. The antenna array is shown as gold rectangles. The array consists of 5 pairs of the two nanorods antennas. The wavelength is 758nm with the gap g=50nm and the pitch along y direction is 253nm. For Fig. 4(a) the dark line is aligned to the left side of the antenna array while for Fig. 4(b), it is aligned to the right. (c),(d) The far field scattering pattern in the XY plane of the arrays corresponding to different alignment of the dark line to the left (c) and to the right (d) of the array.}
\label{fig:false-color}
\end{figure}

Since the antennas are placed near the centre of the singularity point, the scattered power is although detectable, rather weak. A more strongly directional scattering is desired to make the detection easier. To achieve this an array of nanorod pairs can be applied to restrict the scattered power within a very narrow angular range and it has the additional benefit that it also collects a larger part of the incident beam. As an example we apply an array of 5 antennas parallel to the y direction which we illuminate with a Hermite-Gaussian beam with a singular line as shown in Fig. 4(a),(b). The polarization is along the y direction. The pitch along the y direction is chosen as 253nm to achieve maximum lateral scattering based on the array multiplication principle\cite{kraus2002antennas}. The angular scattering pattern largely becomes much narrower compared to the case of a single pair and this further increases the sensitivity of the present scheme as shown in Fig. 4(c),(d). 
		
Finally, we would like to discuss the theoretical detection limit of the present scheme. As shown in the previous discussion, the size of the gap between the two antennas is one important parameter. However, if the gap would be of the order of a few angstroms or less, electrons will start to tunnel from one nanorod to the other\cite{savage2012revealing}, implying that the two antennas would be electrically connected and this prevents the required phase difference to occur. It is worth noting here that focusing of the beam itself is not a necessity, but by focusing the beam, the power can be more concentrated on the nanorods therefore the detected scattering power is increased. In the case of a single nanorod pair, with NA=1.45, gap g=50nm, the scattered far field power at 758nm is approximately 2\% of the total incident power. Using smaller gap makes the observation of the change in the far field pattern more sensitive. 
		
In summary, we have proposed a far field scheme for detection of the deep subwavelength displacement. The scheme is based on the interaction of singular optics with metallic resonant optical antennas which brings together several important areas of research: singular optics, vectorial focal field shaping, resonant nanoscale optical antenna as well as subwavelength nano-metrology into one novel research direction. With the introduction of singularity beam with resonant optical antenna, a new degree of freedom can be brought into the area of detection of deep subwavelength displacement.

、

\begin{acknowledgments}
The authors would like to thank Dr. S.F. Pereira and Prof. J.J.M. Braat for inspiring discussions.

%put your acknowledgments here.
\end{acknowledgments}

%\bibliography{ref}
%

\end{document}